\documentclass{main}

\usepackage{textcomp}
\usepackage{fancyhdr}
\usepackage{lastpage}
\usepackage{amsmath,amssymb}
\usepackage[mathlines]{lineno}

\def\be{\begin{equation}}
\def\ee{\end{equation}}
\def\bea{\begin{eqnarray}}
\def\eea{\end{eqnarray}}

\newcommand{\Photo}{}

\fancypagestyle{firstpagefooter}{%
  \fancyfoot[L]{  \textcopyright \footnotesize This work is an open access article under a Creative Commons Attribution 4.0 International License (https://creativecommons.org/licenses/by/4.0/).}
  \fancyfoot[C]{}
  
}

\begin{document}

\thispagestyle{firstpagefooter}
\title{\Large First measurement of the energy and Mandelstam-$|t|$ dependence of both coherent and incoherent J/$\psi$ photonuclear production
}

\author{\underline{V.~Humlov\'a}\footnote{Speaker, email: vendulka.filova@cern.ch} on behalf of the ALICE Collaboration}

\address{Faculty of Nuclear Sciences and Physical Engineering, Czech Technical University in Prague, Czech Republic}

\maketitle
\abstracts{
A phenomenon, gluon saturation, is expected to emerge in quantum chromodynamics (QCD) at high energies, when gluon splitting and recombination processes reach a dynamic equilibrium. In heavy nuclei, this balance is expected to be achieved at lower energies than in protons, making lead--lead collisions at the LHC an ideal environment to probe the onset of saturation. The diffractive photoproduction of the $J/\psi$ vector meson provides an excellent tool to study this regime, as it directly probes the gluon distribution in the target. ALICE offers unique kinematic coverage of the photon--nucleon centre-of-mass energy, spanning from 20 to 800~GeV, corresponding to three orders of magnitude in Bjorken-$x$ from about $10^{-2}$ down to $10^{-5}$. This contribution presents the latest ALICE results on the energy dependence of coherent $J/\psi$ production, which is sensitive to the average gluon density, and on the energy and Mandelstam-$|t|$ dependence of incoherent production, which probes fluctuations of the gluon field at different spatial scales. These measurements provide new constraints for testing QCD models in the high-energy regime and advance our understanding of the gluonic structure of nuclei.
}

\footnotesize DOI: \url{https://doi.org/xx.yyyyy/nnnnnnnn}


\section{Introduction}
At small Bjorken-$x$ $(x\lesssim10^{-2})$, the partonic structure of hadrons is dominated by gluons whose density increases steeply with energy, as established by deep-inelastic scattering experiments~\cite{H1:2015ubc}. In quantum chromodynamics (QCD), this rapid rise of gluon density is expected to eventually reach a limit at which gluon splitting and recombination processes balance each other, resulting in a dynamical equilibrium known as gluon saturation~\cite{Morreale:2021pnn}. Finding experimental evidence for saturation is among the key challenges of high-energy nuclear physics and a central motivation for ongoing and future collider programs such as the Electron-Ion Collider~\cite{Aschenauer:2017jsk}.

Gluon saturation is predicted to appear at larger values of $x$ (corresponding to lower photon–nucleus energies) in nuclei than in free nucleons due to the high density of color charges in the nuclear environment. The phenomenon can be characterized by the saturation scale $Q_s$, which increases with the nuclear size and energy of the collision. As discussed in Ref.~\cite{Morreale:2021pnn}, saturation arises as a consequence of nonlinear QCD evolution, leading to the formation of a dense regime of gluonic matter in which parton densities no longer grow indefinitely with decreasing $x$. In this picture, ultra-relativistic heavy-ion collisions provide unique access to the nonlinear regime of QCD, particularly through photon-induced processes such as vector-meson photoproduction.

In addition to saturation, parton distribution functions (PDFs) of nucleons bound in nuclei are also modified by coherence effects within the nuclear medium, a phenomenon known as nuclear shadowing~\cite{Armesto:2006ph}. Shadowing results from multiple scattering and quantum interference between partons belonging to different nucleons. It is a leading-twist effect that manifests itself as a depletion of the gluon density in nuclei relative to that of free protons at small $x$.

The relation between shadowing and saturation remains an open question. Both mechanisms lead to a suppression of the nuclear gluon density at small $x$, but they originate from different theoretical frameworks: shadowing arises in the linear regime of QCD evolution, whereas saturation is associated with nonlinear gluon recombination effects. Current data from ultra-peripheral collisions (UPCs) show clear evidence of gluon suppression consistent with both ~\cite{ALICE:2023jgu,Tumasyan:2023PRL}, and distinguishing between them requires precise multi-differential measurements sensitive to spatial and energy scales, such as those provided by coherent and incoherent photoproduction of $J/\psi$ mesons with the ALICE experiment.

The onset of gluon saturation and shadowing effects can be investigated experimentally via diffractive photoproduction of heavy vector mesons, such as the $J/\psi$, in ultra-peripheral collisions.

\section{Diffractive photoproduction of J/$\psi$ in ultra-peripheral collisions}
The electromagnetic field generated by a fast-moving ion with charge $Z$ can be approximated as a flux of quasi-real photons. The intensity of this flux scales as $Z^2$, making heavy-ion collisions a powerful source of high-energy photons. One of these photons can serve as a probe of the target. In lead--lead UPCs, either nucleus can serve as a photon emitter or as a target.

An important process that can occur is diffractive photoproduction, in which an emitted photon fluctuates into a $c\bar c$ dipole which scatters off the target via gluon exchange~\cite{Ryskin:1992ui,Klein:2019qfb}, producing a vector meson. This process is particularly interesting because its cross section is directly sensitive to the gluon dynamics of the target.

Different scenarios can occur depending on the target. If the photon couples coherently to the whole nucleus, the process is referred to as coherent. In such a case, the target nucleus stays intact and only J/$\psi$ is produced. In the case of photon scattering off a single nucleon inside the nucleus, the process is incoherent; the target nucleus breaks up, which usually leads to the emission of forward neutrons in the direction of the incoming target nucleus. The incoherent channel can be further divided into elastic, where the struck nucleon remains intact, and dissociative, in which the nucleon breaks up.

The kinematics of heavy vector meson (J/$\psi$) production, where the hard scale is set by the J/$\psi$ mass, can be reconstructed from the measured rapidity ($y$) and transverse momentum ($p_{\mathrm{T}}$) of the J/$\psi$. The relevant scale of the interaction, usually denoted by $\mu$, is given by the mass of the heavy-quark pair, $\mu^{2} \approx M_{J/\psi}^{2}/4$. This scale defines the resolution at which the gluon density of the target is probed and ensures that the process can be described within perturbative QCD. The relatively large J/$\psi$ mass ($\approx 3.1~\mathrm{GeV}$/$c^2$~\cite{PDG:2024cfk}) therefore provides a hard scale even in photoproduction, where the photon virtuality is very small. The photon--nucleus center-of-mass energy $W_{\gamma \mathrm{Pb,n}}$ is related to the $J/\psi$ rapidity $y$ by
\begin{equation}
W^2_{\gamma \mathrm{Pb,n}} = M_{J/\psi}\sqrt{s_{\mathrm{NN}}}\, e^{\pm y},
\end{equation}
where $M_{J/\psi}$ is the J/$\psi$ mass and $\sqrt{s_{\mathrm{NN}}}$ the nucleon--nucleon center-of-mass energy. The four-momentum transfer squared is approximately $|t|\approx p_T^2$.

Since the photon emitter and target nucleus are indistinguishable in Pb--Pb UPCs, two photonuclear production terms contribute to the observed cross section:
\begin{equation}
\frac{d\sigma_{\mathrm{PbPb}\to J/\psi X}}{dy} = n_\gamma(y)\,\sigma_{\gamma\mathrm{Pb}}(y) + n_\gamma(-y)\,\sigma_{\gamma\mathrm{Pb}}(-y),
\end{equation}
where $n_\gamma$ denotes the photon flux, calculable from the Weizsäcker--Williams approximation~\cite{Broz:2019}, and $\sigma_{\gamma\mathrm{Pb}}$ represents the photonuclear cross section. The measured cross section thus encodes information about both photon directions and therefore about the photon--nucleus center-of-mass energies. To measure the cross section as a function of energy we need to disentangle these contributions.

Moreover, Mandelstam-$|t|$ (with $|t|\simeq p_{\mathrm T}^2$) is related through a Fourier transform to the impact parameter. Within the Good--Walker picture of diffraction~\cite{Good:1960ba}, the coherent and incoherent cross sections can be expressed as the mean and variance, respectively, of the scattering amplitude over all possible configurations of the target. The coherent process, dominated by small momentum transferred ($|t|\lesssim 0.01~\mathrm{GeV}^2$), reflects the average spatial distribution of gluons in the nucleus. The incoherent process, which extends to $|t|$ of order 1~GeV$^2$, is sensitive to fluctuations of the gluon field at nucleon and sub-nucleonic scales. The incoherent cross section is expected to increase with energy until gluon saturation sets in, after which the fluctuations---and hence the cross section---begin to diminish. Measuring both coherent and incoherent production across a wide energy range therefore provides a direct test of this predicted behavior.

\section{Experimental setup and event selection}

 The ALICE detector, in its Run~2 setup,~\cite{ALICE:2014sbx} provides full azimuthal coverage and excellent lepton identification, making it well suited for studies of photon-induced processes.

At forward rapidity ($-4<\eta<-2.5$), $J/\psi\to\mu^+\mu^-$ decays are reconstructed using the Muon Spectrometer, which includes a front hadron absorber, five tracking stations, and a muon trigger system placed behind an iron wall of 7.2 interaction lengths. At midrapidity ($|\eta|<0.9$), $J/\psi\to\mu^+\mu^-$ channel is reconstructed using the Inner Tracking System (ITS) and the Time Projection Chamber (TPC) detectors.

Two arrays of forward scintillators, V0 and AD, provide vetoes against hadronic activity, ensuring the exclusivity of UPC events. The Zero Degree Calorimeters (ZDCs), located 113~m from the ALICE interaction point on both sides, detect forward neutrons emitted in electromagnetic dissociation (for coherent production) or in nuclear breakup following incoherent scattering.

Triggers based on low detector activity and dilepton topologies select exclusive photoproduction events. The classification of neutron emission in the ZDCs is later used to disentangle the photon-emitter and target configurations for both coherent and incoherent processes, as described in the next section.

\section{Disentangling photon directions with the ZDCs}
\label{sec:ZDCclasses}

At forward rapidity, both nuclei can act as photon emitters or targets, producing an ambiguity
between low- and high-energy photon--nucleus interactions. This ambiguity is resolved through the
use of the Zero Degree Calorimeters (ZDCs), which detect forward neutrons on both sides of the
interaction point.

The disentangling mechanism depends on the production process. In the case of coherent photoproduction, additional photon exchanges can lead to electromagnetic dissociation (EMD) of one or both nuclei, producing neutrons independently of the main
interaction. Events are therefore classified according to their neutron multiplicity measured in the ZDCs
into categories (0n0n, 0nXn, XnXn), where X denotes at least one neutron detected in one side of the ZDC. These classes correspond to different average impact parameters and
photon fluxes. By solving a system of differential equations that relate these fluxes to the measured
cross sections, the two photon--target configurations, $\sigma_{\gamma\mathrm{Pb}}(y)$ and
$\sigma_{\gamma\mathrm{Pb}}(-y)$, can be extracted~\cite{Contreras:2017,Baltz:2002pp,Guzey:2013jaa}.

For incoherent photoproduction, the neutron emission originates primarily from the
breakup of the target nucleus caused by the momentum transfer in the hard scattering process.
The direction of neutron emission therefore identifies the photon-emitter nucleus without
additional assumptions: events with neutrons on the same side as the Muon Spectrometer (0nXn) correspond to low-energy photons, while those with neutrons on the opposite side (Xn0n) correspond to high-energy photons. This approach allows for an unambiguous separation of the two photon--target configurations in the incoherent channel.

\section{Measurements of coherent and incoherent J/$\psi$ photoproduction}

The analyses presented here are based on Pb--Pb ultra-peripheral collision data recorded by ALICE in 2018 at $\sqrt{s_{\mathrm{NN}}}=5.02~\mathrm{TeV}$, corresponding to an integrated luminosity of about $533~\mu\mathrm{b}^{-1}$ and $233~\mu\mathrm{b}^{-1}$ for forward and midrapidity measurements, respectively.

\subsection{Energy dependence of coherent production}
\begin{figure}
    \centering
    \includegraphics[width=0.54\linewidth]{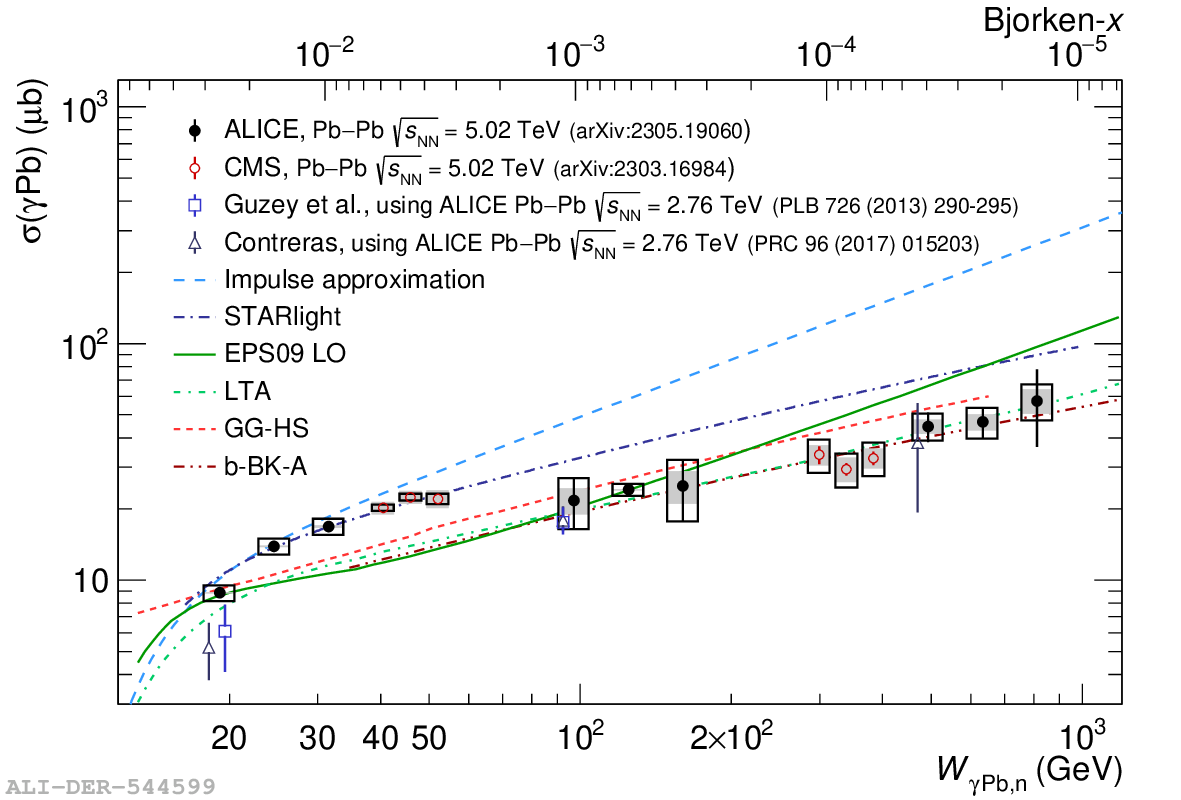}
    \includegraphics[width=0.44\linewidth]{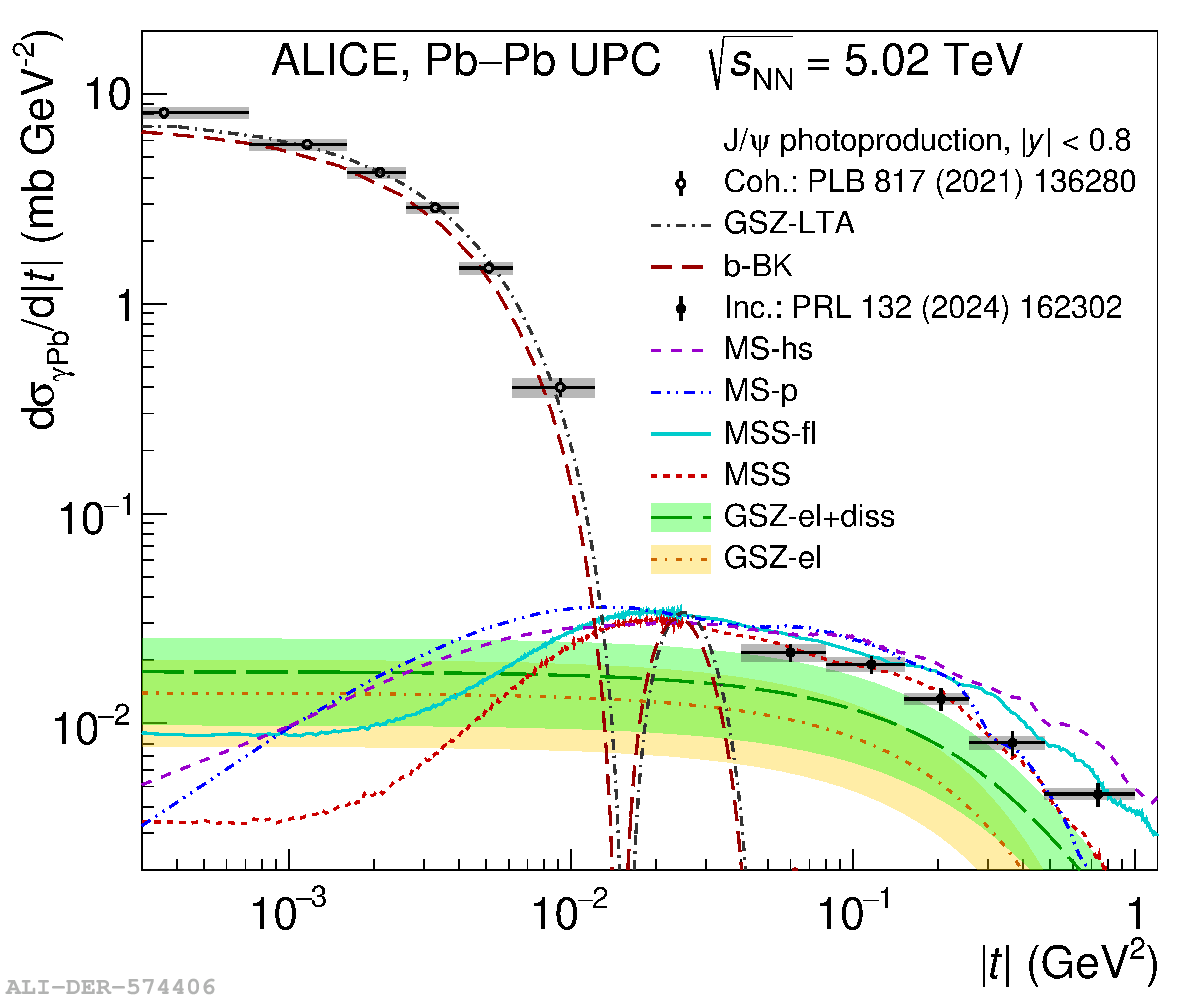}
    \caption{The dependence of the cross section for coherent $J/\psi$ photoproduction on photon-nucleus energy, $W_{\gamma \mathrm{Pb}}$, or Bjorken-$x$ as measured with ALICE~\protect\cite{ALICE:2023jgu} and CMS~\protect\cite{Tumasyan:2023PRL}  (left). Mandelstam-$|t|$ dependence of the cross section for coherent~\protect\cite{ALICE:2021tyx} ($t\lesssim0.01$~GeV$^2$) and incoherent~\protect\cite{ALICE:2023gcs} ($t\gtrsim 0.1$~GeV$^2$) $J/\psi$ photoproduction.}
    \label{fig:Wcoh}
\end{figure}

The coherent $J/\psi$ photoproduction cross section~\cite{ALICE:2023jgu}, measured as a function of the photon--nucleus center-of-mass energy $W_{\gamma \mathrm{Pb}}$, spans from 20 to 800~GeV, corresponding to Bjorken-$x$ values between $10^{-2}$ and $10^{-5}$. The results in Fig.~\ref{fig:Wcoh} (left), show a clear suppression at high energies compared to the Impulse Approximation~\cite{Chew:1952} and STARlight~\cite{Klein:1999} models, which neglect dynamical QCD effects. The data are well described by models incorporating nuclear shadowing within the Leading Twist Approximation (LTA)~\cite{Frankfurt:2011cs} and by saturation-based approaches (b-BK)~\cite{Bendova:2020hbb}. The observed trend indicates a reduction of the effective gluon density in the nucleus at small $x$, consistent with shadowing and/or saturation effects. Moreover, the measurement performed by CMS~\cite{Tumasyan:2023PRL} complements the Bjorken-$x$ coverage of ALICE, and the results from both experiments follow the same trend.

\subsection{Mandelstam-$|t|$ dependence}
At midrapidity, the $|t|$-dependence of the coherent and incoherent photonuclear $J/\psi$ cross sections, measured at $\sqrt{s_{\mathrm{NN}}}=5.02$~TeV~\cite{ALICE:2021tyx,ALICE:2023gcs}, reveals distinct features of the nuclear structure, see Fig.~\ref{fig:Wcoh}, right. The coherent~\cite{ALICE:2021tyx} process dominates at $|t|<0.01$~GeV$^2$, where the photon couples to the entire nucleus, and its slope reflects the transverse size of the nuclear gluon distribution. The data favor shadowing-based (LTA~\cite{PhysRevC.95.025204}) and saturation-based (b-BK~\cite{PhysRevD.100.054015}) models. The incoherent~\cite{ALICE:2023gcs} component extends up to $|t|\sim 1$~GeV$^2$, probing fluctuations at the nucleon and sub-nucleonic scales. Models including subnucleonic degrees of freedom reproduce the data more accurately at large $|t|$ than those assuming smooth nuclear profiles~\cite{Mantysaari:2022sux,Cepila:2023dxn}.

\subsection{First measurement of both energy and Mandelstam-$|t|$ dependence of incoherent production}

The recent ALICE analysis presents the first measurement of the incoherent $J/\psi$ photoproduction cross section simultaneously as a function of the photon--nucleus energy $W_{\gamma \mathrm{Pb,n}}$ and the four-momentum transfer squared $|t|$ in Pb--Pb ultra-peripheral collisions at $\sqrt{s_{\mathrm{NN}}} = 5.02~\mathrm{TeV}$~\cite{ALICE:2025cuw}. 
The study extends the previous separate measurements of the energy~\cite{ALICE:2023jgu} and $|t|$~\cite{ALICE:2021tyx,ALICE:2023gcs} dependences and provides new insight into the evolution of nuclear gluon distributions with both energy and spatial scale in transverse plane.

J/$\psi$ mesons were reconstructed in the dimuon decay channel using the Muon Spectrometer ($-4 < \eta < -2.5$). 
Forward neutron emission measured by the ZDCs was used to tag the photon direction, thus identifying the photon emitter and target nuclei and providing access to both high ($W_{\gamma \mathrm{Pb,n}} \sim 600~\mathrm{GeV}$) and low ($W_{\gamma \mathrm{Pb,n}} \sim 20$--$30~\mathrm{GeV}$) energy configurations.

The incoherent cross section was determined in three $|t|$ intervals ($0.09 < |t| < 0.36$, $0.36 < |t| < 0.81$, and $0.81 < |t| < 1.44~\mathrm{GeV}^2$) by fitting the dimuon invariant-mass distributions with a double-sided Crystal Ball function for the signal and an exponential for the background. 
The contamination from coherent production and from $\psi(2\mathrm{S})$ feed-down was estimated through template fits to the $p_{\mathrm{T}}$ distributions using STARlight simulations~\cite{Klein:2016yzr} and the H1 parameterization of nucleon-dissociative events~\cite{H1:2013okq}. 
Systematic uncertainties were dominated by the muon trigger efficiency, number of V0C cells fired, and signal extraction.

\begin{figure}
    \centering
    \includegraphics[width=0.45\linewidth]{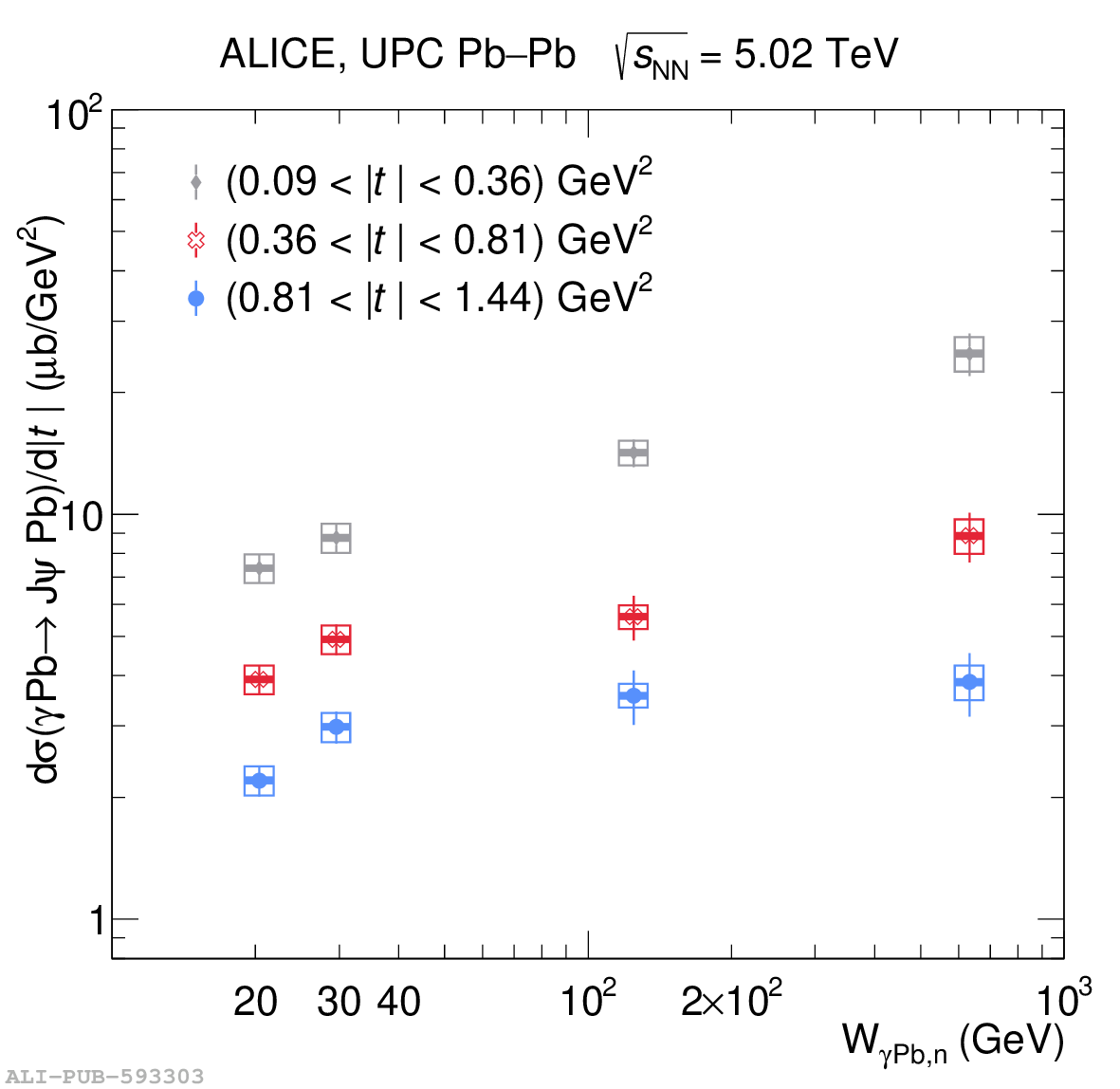}
    \includegraphics[width=0.45\linewidth]{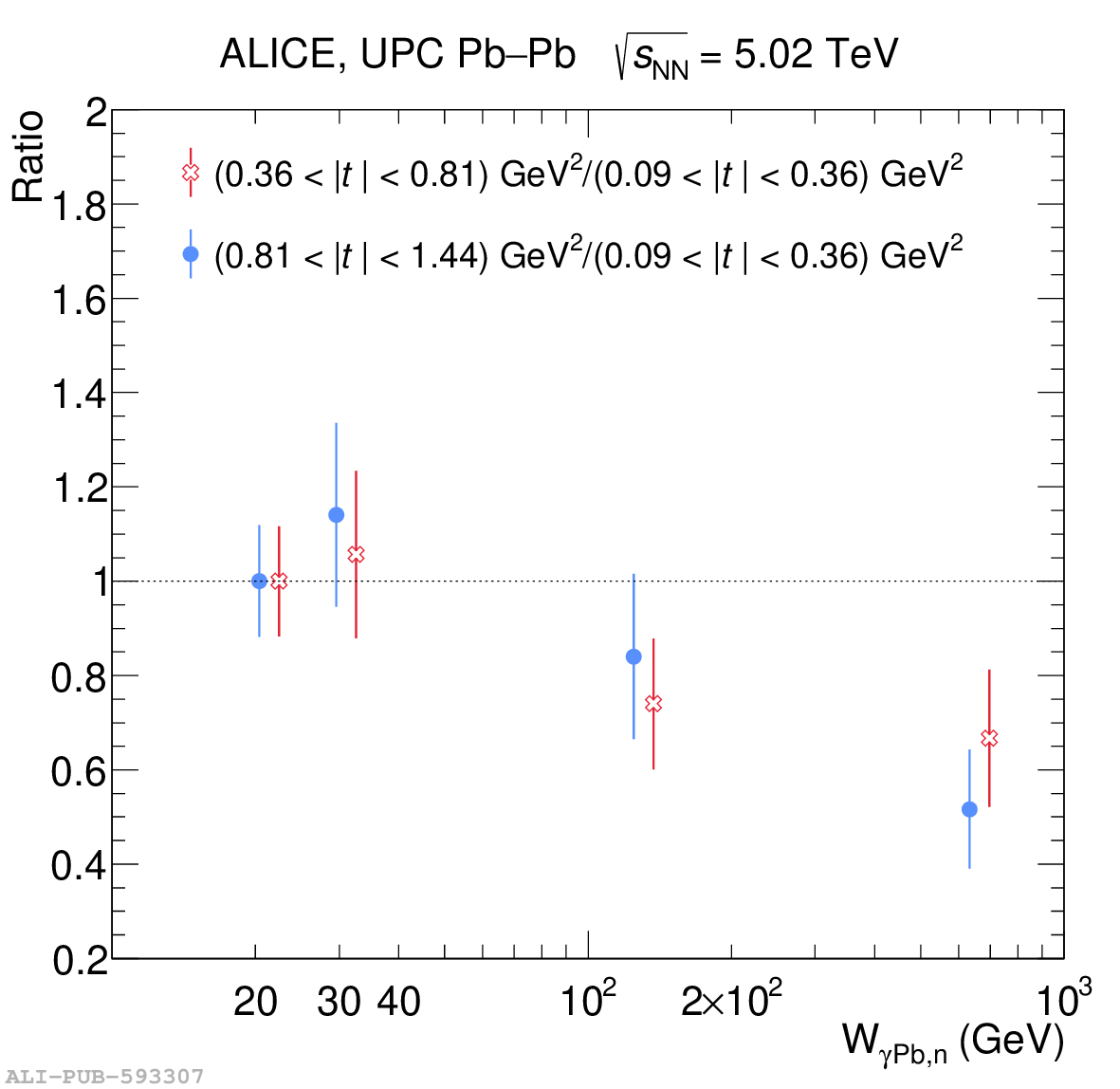}
    \caption{The energy dependence of the cross section of incoherent $J/\psi$ photonuclear production off lead nuclei is shown for three different ranges of the Mandelstam-$|t|$ variable~\protect\cite{ALICE:2025cuw} (left). The symbols denote the measured cross section, the vertical lines across them represent the uncorrelated uncertainty, the open boxes the correlated uncertainty, and the horizontal shadow box the uncertainty from the computation of the photon flux. Ratios of the cross sections at different Mandelstam-$|t|$ ranges~\protect\cite{ALICE:2025cuw} normalized such that the lower energy ratio is one. The vertical line denotes the statistical and $|t|$-uncorrelated systematic uncertainties added in quadrature.}
    \label{fig:gammaPbW}
\end{figure}

The left plot of Fig.~\ref{fig:gammaPbW} shows the measured differential cross sections $\mathrm{d}\sigma_{\gamma \mathrm{Pb}}/\mathrm{d}|t|$ as a function of $W_{\gamma \mathrm{Pb,n}}$ for the three $|t|$ intervals. 
The cross section increases with energy in all intervals, but the rate of this growth is suppressed at larger $|t|$ w.r.t. small $|t|$, where smaller gluonic configurations are probed. 
This suppression indicates a reduction of event-by-event fluctuations of the gluon field.

To quantify this behavior, ratios of the cross sections at large to small $|t|$ were evaluated as a function of $W_{\gamma \mathrm{Pb,n}}$, normalized to unity at low energy. 
At $W_{\gamma \mathrm{Pb,n}} = 633~\mathrm{GeV}$, the ratio between the highest and lowest $|t|$ intervals reaches $0.52 \pm 0.13$, deviating from unity by more than three standard deviations, see right panel in Fig.~\ref{fig:gammaPbW}. 
This represents the first experimental evidence that the growth of the incoherent cross section with energy is reduced when probing subnucleonic spatial scales.

\begin{figure}
    \centering
    \includegraphics[width=.99\linewidth]{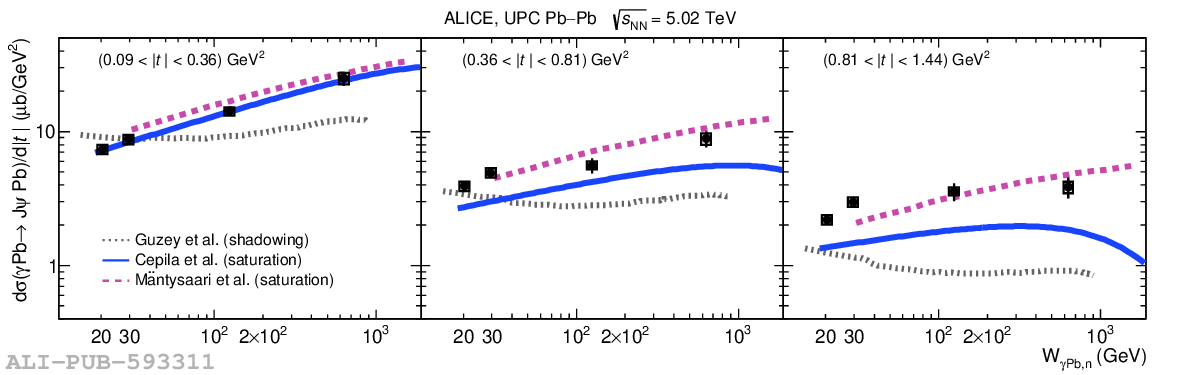}
    \caption{The energy dependence of the cross section of incoherent J/$\psi$ photonuclear production at three ranges of the Mandelstam-$|t|$ compared with model predictions~\protect\cite{ALICE:2025cuw}. The shadowing-based model of Guzey et al.~\protect\cite{Guzey:2018tlk} is shown with a dot-dashed line. Two saturation-based predictions are also shown: from M\"antysaari et al.~\protect\cite{Mantysaari:2022sux,Mantysaari:2023xcu} with a dashed line and from Cepila et al.~\protect\cite{Cepila:2023dxn} with a solid line.}
    \label{fig:c3in1}
\end{figure}

The data were compared with theoretical calculations that incorporate different physical mechanisms, as shown in Fig.~\ref{fig:c3in1}. 
The shadowing-based model by Guzey et al.~\cite{Guzey:2018tlk}, which successfully reproduces coherent $J/\psi$ production~\cite{ALICE:2023jgu}, is disfavoured by the incoherent cross section data. 
In contrast, saturation-based models that include subnucleonic fluctuations provide a better description. 
The energy-dependent hot-spot model of Cepila et al.~\cite{Cepila:2023dxn} attributes the observed suppression to the overlap of an increasing number of gluonic hot spots at high energy, reducing the variance of the gluon density. 
The framework of Mäntysaari et al.~\cite{Mantysaari:2023xcu} also reproduces the data trend through a decreasing power-law exponent with increasing $|t|$, reflecting a slower growth of the cross section at small transverse scales. 
Both approaches qualitatively match the measurements, with the model of Mäntysaari et al. providing the closest overall agreement.

\section{Conclusion and outlook}
The ALICE measurements presented here provide a comprehensive view of diffractive $J/\psi$ photoproduction in ultra-peripheral Pb--Pb collisions, linking the energy and momentum-transfer dependencies of both coherent and incoherent processes. The coherent results show a strong suppression of the cross section at high photon--nucleus energies, consistent with nuclear shadowing and gluon saturation effects at small Bjorken-$x$. At midrapidity, the $|t|$-dependence of the incoherent channel reveals a shallower slope than predicted by nucleon-level models, indicating fluctuations of the gluon density at subnucleonic scales.

The new multi-differential measurement of the incoherent cross section as a function of both $W_{\gamma \mathrm{Pb,n}}$ and $|t|$ shows that the growth of the cross section with energy becomes weaker at large $|t|$, where smaller gluonic configurations are probed. This behaviour suggests a change in the fluctuation pattern of the gluon density with decreasing transverse scale. Incoherent photoproduction thus provides a sensitive probe of the spatial structure and its fluctuations in nuclei at small Bjorken-$x$.

With the increased luminosity and upgraded detectors in Runs~3 and~4, ALICE will extend these studies to lower Bjorken-$x$ and finer $|t|$ intervals, enabling a more detailed mapping of the nuclear gluon density and its fluctuations.


\end{document}